\relax
\documentclass[letterpaper]{article} 
\usepackage{paralist}
\usepackage{aaai212}  
\usepackage{times}  
\usepackage{helvet} 
\usepackage{courier}  
\usepackage[hyphens]{url}  
\usepackage{graphicx} 
\usepackage{natbib}  
\usepackage{caption} 
\usepackage{booktabs}
\usepackage{xspace}

\newcommand{\descr}[1]{\vspace{0.07cm}\noindent\textbf{#1}}

\title{The Evolution of the Manosphere Across the Web\thanks{\small This paper has been accepted at the 15th International Conference on Web and Social Media (ICWSM), please cite accordingly.}}
\author{
Manoel Horta Ribeiro,$^{\spadesuit, *}$
Jeremy Blackburn,$^{\triangle}$
Barry Bradlyn,$^{\diamondsuit}$
Emiliano De Cristofaro,$^{\nabla}$\\
Gianluca Stringhini,$^{\clubsuit}$ 
Summer Long,$^{\diamondsuit}$
Stephanie Greenberg,$^{\diamondsuit}$
Savvas Zannettou$^{\heartsuit, *}$\\}

\affiliations{
$\spadesuit$ EPFL,
$\triangle$ Binghamton University,
$\diamondsuit$ University of Illinois at Urbana-Champaign \\
$\nabla$ University College London,
$\clubsuit$ Boston University,
$\heartsuit$ Max Planck Institute for Informatics\\
$*$ Corresponding authors: manoel.hortaribeiro@epfl.ch, szannett@mpi-inf.mpg.de
}

\begin{document}

\maketitle

\begin{abstract}
We present a large-scale characterization of the Manosphere, a conglomerate of  Web-based misogynist movements focused on ``men's issues,'' which has prospered online.
Analyzing 28.8M posts from 6 forums and 51 subreddits, we paint a comprehensive picture of its evolution across the Web, showing the links between its different communities over the years.
We find that milder and older communities, such as Pick Up Artists and Men's Rights Activists, are giving way to more extreme ones like Incels and Men Going Their Own Way, with a substantial migration of active users.
Moreover, our analysis suggests that these newer communities are more toxic and misogynistic than the older ones.
\end{abstract}

\section{Introduction}

Online platforms are increasingly exploited to spread hate, extreme ideologies, and weaponized information, and have been repeatedly linked to radicalization leading to real-world violent events~\cite{allison2018social,kowalski2013psychological}.
Seemingly niche communities are often involved in such activities; for instance, Gab, 8chan, and 4chan have all played a role in the apparent radicalization of individuals that went on to alleged murderous actions~\cite{christchurch,4chan-bellingcat}. 
Each of these communities was considered niche;
yet, \emph{niche} does not mean \emph{unimportant}.
In fact, these are often well-positioned within online cultures to have impactful effects in the world~\cite{nagleKillAllNormies2017}.

Over the past few years, the \emph{Manosphere} has emerged as a noteworthy conglomerate of  ``niche'' communities, roughly aligned by their common interest in masculinity and its alleged crisis~\cite{lillyWorldNotSafe2016}.
Communities like Pick Up Artists (PUAs), Men's Rights Activists (MRAs), Men Going Their Own Way (MGTOW), and Involuntary Celibates (Incels) have been growing in size and in their involvement in online harassment and real-world violence~\cite{waposb}.

So far, researchers have mostly studied the Manosphere through deep domain experience and immersion into the material, drawing from a tradition of diverse concepts, such as techno-sociological theories (e.g., toxic disinhibition) and views on patriarchal societal constructs~\cite{lumsden2019WantKillYou,gingAlphasBetasIncels2017,linAntifeminismOnlineMGTOW2017,farrell2019exploring}.

However, Manosphere communities are scattered through the Web in a loosely connected network of subreddits, blogs, YouTube channels, and forums~\cite{lewisLearnFarRight2019}.
Consequently, we still lack a comprehensive understanding of the underlying digital ecosystem, of the evolution of the different communities, and of the interactions among them.

\descr{Present Work.} In this paper, we present a multi-platform longitudinal study
of the Manosphere on the Web, aiming to address three main
research questions:

\begin{enumerate}
     \item[\textbf{RQ1:}] How has the popularity/levels of activity of the different Manosphere communities evolved over time? 
     \item[\textbf{RQ2:}] Has there been substantial migration of, or intersection in, users across communities?
     \item[\textbf{RQ3:}] Has speech become more toxic and/or misogynistic over time?
\end{enumerate}

To answer these questions, we collect and analyze a large dataset with posts collected from 6 forums (6.7M posts) and 51 subreddits (22.1M posts) related to the Manosphere.
We annotate and categorize each forum/subreddit according to the most relevant movement (\emph{e.g.}, Incels, MGTOW, PUA), and study their evolution over time.
Specifically, we:
1) analyze the evolution of the user base and of the activity in each of the different subreddits and the forums over time;
2) study the user migration flow across communities by examining, in different subreddits, the intersection between users in different communities;  and
3) examine the toxicity and the misogynistic content in the communities across the years using the Perspective API~\cite{jigsaw2018perspective} and a misogyny-related lexicon~\cite{farrell2019exploring}.

We find that older communities such as Men's Rights Activists and Pick Up Artists have eventually been overshadowed by new communities like Men Going Their Own Way, and Incels (\textbf{RQ1}).
Moreover, users are often active across several Manosphere communities, with newer communities receiving significant migratory influxes from the older ones (\textbf{RQ2}).
Newer communities are also appreciably more toxic and misogynistic, which might imply a trend of the Manosphere as a whole (\textbf{RQ3}).
Overall, our work highlights the importance of studying different communities at scale, in aggregate, rather than separately, and prompts a number of possible future directions, including understanding the reasons behind user migration movements. Code and data are available at \url{www.doi.org/10.5281/zenodo.4007913}.

\section{Background and Related Work}\label{sec:background}

The Manosphere is loosely defined as a collection of communities aligned by their common interest in men's issues, often associated with
online harassment and real-world violence~\cite{waposb}. 
Since the creation of the term around 2009~\cite{Manosphereblog}, it has been adopted by the media and by those within the Manosphere~\cite{ironwoodManospherebook}. 

\descr{Origins of the Manosphere.}
The ``roots'' of the Manosphere can be traced back to the Men's Liberation Movement in the 60s and 70s. 
The movement was a critique of traditional male gender roles, which were seen as oppressive~\cite{messner1998limits}. 
In the 70s, a new branch began to see the ``problem''
as stemming more from feminism and women empowerment~\cite{messner1998limits}.
So-called Men's Rights Activists would focus on men's issues such as military conscription, divorce, and custody laws. 
According to this new ideology, women's liberation would be inflicting on men ``the worst of both worlds,'' and the movement's empathetic tone
turned into anger~\cite{costonWhiteMenNew2012}.

Some of the lead figures of these movements were once associated with second-wave feminism; 
Warren Farrell, for example, led a men's group within the National Organization for Women. 
Then, in 1993, he wrote the Myth Of Male Power~\cite{farrell1996myth}, which became a fundamental text to Men's Rights Activists, claiming that \emph{men}, and not women, 
are systematically disadvantaged in society.
Farrell's work has been criticized as simplistic and insensitive to women's struggle for equality~\cite{mifflinmargotMythMalePower1993EntertainmentWeekly,beasley2005gender}.

\descr{Manosphere Across the Web.} 
\citet{gingAlphasBetasIncels2017} argues that many traditional theories on views of both masculinity and misogyny are expressed in new, impactful ways online.
As modern society has generally accepted the core ideals of gender equality, she observes an increase in ``a more virulent strain of anti-feminism online,'' positing that the Web's facilitation of information dissemination across boundaries and platforms has increased the spread of extreme anti-feminist views, along with misogyny and violent rhetoric.

\citet{farrell2019exploring}
present a large-scale analysis of Manosphere-related subreddits with a focus on misogyny.\footnote{Note that this work is by Tracy Farrell, and not Warren Farrell.}
They build a lexicon from seven dictionaries related to harassment, violence, etc., and manually categorize the terms in this meta-lexicon into one of nine misogynistic categories informed from feminist theory (e.g., ``belittling,'' ``patriarchy,'' ``physical violence,'' etc.).
Out of 2,454 terms, they select 1,300 for their final lexicon, which is then used to measure the prevalence of terms from each category in 5.7M posts across six  Manosphere-related subreddits.

\citet{laviolette2019using} study two subreddits related to the Manosphere (\emph{/r/MensRight} and \emph{/r/MensLib}).
Using machine learning and qualitative analysis, they characterize both subreddits with a particular focus on understanding their interpretations regarding gender and discrimination.
 
\citet{lillyWorldNotSafe2016} offers an overview of the representational politics of the Manosphere.
Through mixed-methods critical discourse analysis, she presents a taxonomy for the communities and highlights the key ideas of the Manosphere: 
masculinity is under siege by feminizing forces; 
and feminism is hypocritical and oppressive. 
Her taxonomy considers 4 communities, Men's Rights Activists (MRA), Men Going Their Own Way (MGTOW), Pick Up Artists (PUA), and Involuntary Celibates (Incels).
According to her, while the first two communities see this ``crisis in masculinity'' in society, which would be increasingly ``feminized,'' the two latter see the crisis in men themselves.

\descr{Manosphere Taxonomy.} We describe the communities of the Manosphere as defined by \citep{lillyWorldNotSafe2016}'s taxonomy:

\begin{enumerate}
\item \textbf{Men's Rights Activists (MRA)} focus on men-related social issues and institutions, which they argue discriminate against men~\cite{costonWhiteMenNew2012}. 
The movement has been repeatedly labeled as misogynistic and/or hateful~\cite{maddisonPrivateMenPublic1999,goldwagLeaderSuicideBrings2012}.
\item \textbf{Men Going Their Own Way (MGTOW)} also believe society is rigged against men~\cite{linAntifeminismOnlineMGTOW2017}. They espouse the abandonment of women and sometimes, of western society. The system is impossible to change, so the solution is to ``go your own way.''
This is often paired with extreme anti-feminism and misogyny~\cite{smithWhyTheseStraight2016}.
\item \textbf{Pick Up Artists (PUA)} are a community built around ``game:'' techniques, strategies, and mindsets that help men pick up women~\cite{lillyWorldNotSafe2016}. 
This often involves objectifying women and promoting harassment techniques like \textit{negging}~\cite{waposb}, \emph{i.e.}, insulting women to undermine their confidence. 
The PUA community conceptualizes the masculinity crisis in terms of the femininization of the man, ``a fool at the hands of women''~\cite{lillyWorldNotSafe2016}. 
\item \textbf{Involuntary Celibates (Incels)} are a group, mostly of young men, 
united by a strong feeling of rejection and rage towards the opposite sex. 
Incels rose to the mainstream due to their association with mass murderers~\cite{HowRampageKiller2018}. 
The community is obsessed with theories about looks and relationships, and members often express a desire to hurt others or themselves. 
\end{enumerate}

\descr{Hateful and Abusive Speech.}
Prior work has studied how to detect and measure hate speech on the Web.
\citet{silva2016analyzing} analyze the targets of hate speech in Whisper and Twitter by matching expressions of hate towards something.
\citet{chandrasekharan2017you} examine 100M posts and comments from two banned subreddits, \emph{/r/fatpeoplehate} and \emph{/r/CoonTown}, and observe user behavior on other subreddits after the ban by the number of occurrences of a set of keywords.
\citet{Olteanu2018} measure the causal impact of extremist violence on hate speech online, also using a lexicon-based approach.

Detecting and measuring hateful content is hard, as hate speech is contextual and subjective~\cite{sellars2016defining,schmidt2017survey,Ribeiro2018}.
Recent work highlights these challenges empirically: 
\citet{Davidson2017} shows that machine learning algorithms struggle to distinguish offensive and hateful speech,
while~\citet{hosseini2017deceiving} points to possible ways to deceive APIs.
\citet{davidson_racial_2019} find that techniques for hate speech detection tend to predict tweets written in African-American English as abusive at a substantially higher rate.

In this paper, we focus on measuring toxic content and misogynistic speech using two methods: 1) Google's Perspective API \cite{jigsaw2018perspective}; 
and 2) a dictionary-based approach relying on misogyny-related keywords from \cite{farrell2019exploring}.
Naturally, this shares the caveats of the aforementioned work, which, however, we mitigate by using two {\em different} methods and by comparing their signal across time in multiple communities and platforms.

\descr{Fringe communities.} 
Prior work has also studied {\em fringe} communities.
\citet{bernstein20114chan} analyze 5M posts on 4chan's random (/b/) board to examine the effect of anonymity and ephemerality, while \citet{hine2017kek} focus on the politically incorrect board (/pol/), studying 8M posts collected over almost three months.
\citet{Zannettou2018} study the influence of fringe communities with respect to images.
\citet{ribeiro2020auditing} study the growth of the Alt-right on YouTube through ``radicalization pathways.''
Finally, anecdotal evidence points towards a significant intersection between the Manosphere and other fringe communities~\cite{futrelle_alt_2020}.

\section{Datasets}
We use data we collected from two main sources:
1) six forums related to the Manosphere; 
and
2) posts from 51 subreddits related to the Manosphere.

\descr{Forums.} 
Forums are an interesting medium through which to study the Manosphere, as they allow us to track communities that were prominent before Reddit (e.g., PUAs).
Moreover, they often involve discourse considered too toxic for mainstream social media platforms~\cite{nashrullaIncelsAreRunning2019BuzzFeedNews}.
Thus, we crawl multiple forums related to four Manosphere-related communities: Incels, PUA, MRA, and MGTOW, based on manual visits of specialized Wikis -- specifically, IncelsWiki, RationalWiki, and RedPillWiki.\footnote{Sources: \url{incels.wiki}, \url{incels.wiki/w/incels.co}, \url{rationalwiki.org}, and \url{redpilltalk.com}.}

We focus on forums that: 
1)~are popular and/or active for a long period of time;
2)~do not require registration to see all posts; and
3)~are broadly related to the communities of interest.
We select six forums\footnote{Sources: \url{incels.is}, \url{rooshvforum.com}, \url{pick-up-artist-forum.com}, \url{theattractionforums.com}, 
\url{mgtow.com/forums}, \url{forums.avoiceformen.com}.}, described below.

\begin{compactenum}

\item \textbf{Incels.is:} Forum created hours after the ban of \emph{/r/Incels}~\cite{nashrullaIncelsAreRunning2019BuzzFeedNews}.  It is the largest Incel forum in activity, allowing speech that would would likely be censored in platforms like Reddit~\cite{incelwikiIncelsCo2019}.  It does not allow women to join as they would disrupt discussion~\cite{badgerlivestreamsTalkingIncelsCo2019}.
\item \textbf{Rooshv:} Created by prominent ex-PUA Daryush V. It was a place to discuss game,
and sex-tourism, but it increasingly became fertile ground for anti-feminist and Alt-right narratives~\cite{wyattSexTouristSwallows2013}. The discourse changed so much that in 2019, Daryush condemned extramarital sex as sinful~\cite{shugermanNoxiousPickupArtist2019}.

\item \textbf{The Attraction and MPUA Forum:}
Older forums dedicated to pick-up-artistry.
The Attraction is run by \url{lovesystems.com},
and MPUA Forum featured coaches~\cite{mpuaforumsGambler2019} and sold courses/event tickets.

\item \textbf{MGTOW Forum:} The largest MGTOW forum, created in 2014. Threads often discuss alleged negative female traits, society's supposed bias against men, and how men should abandon women altogether~\cite{rationalwikiMenGoingTheir2019}.

\item \textbf{AVFM:} Official forum from Paul Elam's MRA site \textit{A Voice for Men}~\cite{caseyWeWentMen2013}. Created in 2012, it was a space to discuss topics such as anti-feminism, divorce, and reproductive choice.

\end{compactenum}
For each forum, we built a custom crawler, which we ran between June 19--30, 2019.
Importantly, all content in forums was publicly available and did not require us to log in.
Table~\ref{tab:dataset_overview} (top part) lists the number of threads, users, and posts, as well as the relevant dates.
Overall, we gather 6.7M posts from 130K users.

\begin{table}[t]
{
\footnotesize
\begin{center}
\footnotesize

\begin{tabular}{lrrr}
\toprule
\textbf{Forum} &  \#\textbf{Threads} &  \#\textbf{Users} &     \#\textbf{Posts}  \\
\midrule
Incels.is &    117,592 &    5,933 &   2,436,053  \\
Rooshv &     47,203 &   13,598 &   1,490,231  \\
The Attraction &    133,520 &   45,937 &     892,715 \\
MPUA Forum &    165,270 &   59,332 &     890,651  \\
MGTOW Forum &     52,022 &    3,762 &     794,698  \\
  AVFM &     20,159 &    1,804 &     211,070  \\ \midrule
 {\bf Total} &  535,766	& 130,366	& 6,715,418 \\\toprule
\textbf{Subreddits} &  \#\textbf{Threads} &  \#\textbf{Users} &     \#\textbf{Posts} \\ \midrule
PUA (7 subr.)&          228,208 &  177,095 &  2,041,313 \\
Incels (18) &           419,819 &  197,196 &  6,413,459   \\
TRP (12)&         272,673 &  162,192 &  5,286,137   \\
MRA (11)&         241,053 &  214,061 &  4,691,807  \\
MGTOW (3)\hspace*{-0.5cm}&           199,702 &   85,268 &  3,728,367  \\\midrule
 {\bf Total} &  1,361,455 & 835,812 &  22,161,083  \\
   
\bottomrule

\end{tabular}
\end{center}
\caption{Overview of our dataset (not including baselines).}
\label{tab:dataset_overview}
}
\end{table}


\descr{Reddit.} 
We first retrieve all submissions and comments made available via Pushshift between June 2005 and December 2018~\cite{baumgartner2020pushshift}.
Then, we identify a set of subreddits relevant to the Manosphere by finding references to subreddits on the Incels Wiki page as well as popular subreddits like \emph{/r/MGTOW}, \emph{/r/Braincels}, \emph{/r/TheRedPill}.
In the end, we select 56 subreddits.

By parsing Pushshift's monthly dumps, we extract all submissions and comments for each of the subreddits.
Note that we find data for 51 out of 56 subreddits, as five subreddits were banned before Pushshift collected data for them.
Ultimately, we gather a set of 22M posts from 835K users.
Next, we group the subreddits as into five categories, largely based on the taxonomy of Manosphere subcultures~\cite{lillyWorldNotSafe2016}:

\begin{compactenum}
\item \textbf{\em Incels:} discusses involuntary celibacy or related concepts, e.g., the Black Pill.       
\item \textbf{\em MGTOW:} discusses the MGTOW movement.        
\item \textbf{\em PUA:} discusses pick up artistry and game.      \item \textbf{\em MRA:} discusses the MRA movement and its broader agenda (e.g., anti-circumcision, alimony).       
\item \textbf{\em TRP:} associated with \emph{/r/TheRedPill}, as well as other subreddits that broadly belong to the Manosphere.
\end{compactenum}

While the first 4 categories {\em directly} map to Lily's taxonomy, we separate the subreddits related to \emph{/r/TheRedPill} into a different category.
Journalistic investigations~\cite{romano_reddits_2017} suggest that the \emph{/r/TheRedPill}'s creator, pk\_atheist, was a state lawmaker with former ties to the Men's Rights movement.
However, the subreddit (as well as other ``adjacent'' ones like \emph{/r/RedPillParenting} and \emph{/r/RedPillWomen}) merges elements from several Manosphere communities.
For instance, it has a lot of content related to relationships and pick-up artistry~\cite{tait_spitting_2017}\footnote{\url{https://incels.wiki/w/R/TRP}}, but many of the members self-identify as MGTOW, to the point where they have a flair (a category of posts) dedicated to MGTOW-related content.\footnote{See, e.g. \url{www.reddit.com/r/TheRedPill/comments/2sd0cg/}} 

\descr{Annotation.} 
To assign each subreddit to one of the five categories, we performed manual annotation with seven annotators. 
Each annotator spent a minimum of ten minutes browsing the subreddits and reading random posts from them and assigned each subreddit to one category.
The subreddits were then labeled based on majority agreement.
Note that the annotators are authors of this paper familiar with the scholarly literature on the Manosphere. Also, they had no communication with each other about the annotation task.
We compute the Fleiss' Kappa Score to assess the agreement between the annotators, which amounts to 0.91; this is regarded as ``almost perfect agreement,'' hence denoting high inter-annotator agreement~\cite{landis1977measurement}. 
In Table~\ref{tab:dataset_overview} (bottom part), we provide the number of subreddits, threads, users, and posts, for each of the categories. 

\descr{Baseline Datasets.} 
Beyond the subreddits and forums already discussed, we use a few additional data sources as baselines.
Namely, we use a dataset from Gab~\cite{zannettou2018gab} consisting of 29M posts, using the publicly available corpus\footnote{Available at \url{https://files.pushshift.io/gab/}}, a random dataset consisting of 0.5\% of all the Reddit corpus, as well as a dataset from 4chan's Politically Incorrect board (/pol/).
To extract the random dataset from Reddit, we parse all posts between 2005-2019 and generate a random sample of 0.5\% of all posts (28M posts).
For 4chan's /pol/, we select 25\% of all posts (32M posts) from the dataset collected by \cite{hine2017kek}, with all posts from June 2016 to September 2019.
These are employed through the paper as reference points for our other analyses.

\descr{Ethics.}
We follow standard ethical guidelines~\cite{rivers2014ethical}, not making any attempt to de-anonymize or link users across platforms, encrypting data at rest, etc.
We work exclusively with publicly available data.

\begin{figure*}[t]
\center
\includegraphics[width=\textwidth]{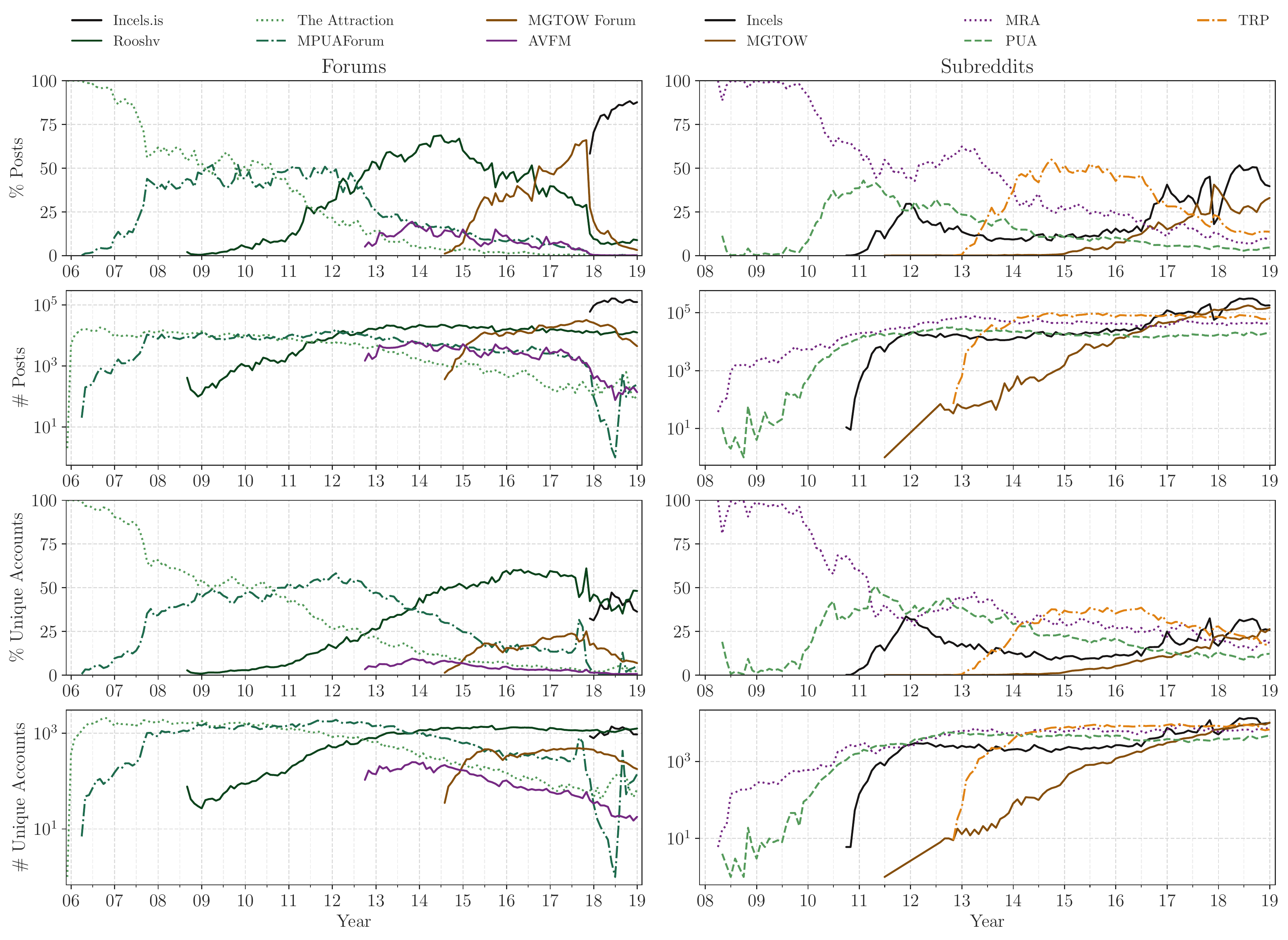}
\caption{Evolution of the number of posts and active accounts at each point in time in the forums (left-hand side) and subreddits (right-hand side). We depict, for each month, both the normalized and the absolute number of posts (top two rows), and active accounts (bottom two rows). 
The normalization is done separately for forums and subreddits, and thus in the first and third rows, the percentages always sum to 100. 
}
\label{fig:all}
\end{figure*}

\section{Evolution of the Popularity and of Activity Levels in the Manosphere (RQ1)}

To study the changes in popularity and activity in the different Manosphere communities over time, we longitudinally examine the number of active users and posts.
We present these quantities aggregated per month both in absolute terms and normalized over the total number of active accounts.\footnote{We normalize over all monthly active accounts/posts in all forums/subreddits.}
The former gives us a sense of the absolute growth of the Manosphere and of each community, while the latter allows us to better compare the changes in the relative sizes of the communities.
We consider users to be \emph{active} in a given month if they have posted at least once.
For forums, as we discuss their timelines, we use the Internet Archive to check whether the first posts are indeed indicative of the beginning of the community.
Fig.~\ref{fig:all} depicts, in the top two rows, the normalized percentage and the number of monthly posts for forums (left) and subreddits (right).
Below, in the bottom two rows, we show the same values for  monthly active accounts.
We discuss the trends in popularity and activity of each community separately.

\descr{Pick Up Artists.} 
PUA-related forums (\emph{The Attraction}, \emph{MPUAForum}, \emph{Rooshv}) were the first to be created and quickly grew in number of active monthly accounts and posts.
The first two forums, by the end of 2008, had over a thousand active monthly users and had reached close to their maximum number of monthly posts.
This coincides with a time where PUAs got a lot of attention in the media. 
In September 2005, Neil Strauss wrote ``The Game''~\cite{strauss_game_2015} an expos\'e of the PUA community that was featured as a New York Times Bestseller, and between 2007 and 2008 the television show ``The Pick Up Artist'' aired on VH1.

The first posts from \textit{Rooshv} date back to mid-2008, which is earlier than its first registry in the Internet Archive.\footnote{\url{web.archive.org/web/*/www.rooshvforum.com}}
This was around the same time the first PUA subreddits were created, \emph{/r/PUA} and \emph{/r/seduction}, which are still the largest.
Unlike \emph{The Attraction} and \emph{MPUAForum}, which declined in number of active accounts and posts in the following years, \emph{Rooshv} and the PUA-related subreddits continued to prosper, and present a relatively stable number of posts since 2013.
It is worth recalling that as it evolved, Rooshv morphed into a very different community, more aligned with the Alt-right~\cite{shugermanNoxiousPickupArtist2019}.

Looking at the normalized percentage of posts and active accounts, we find that PUA forums had a large share of the total number of posts and active accounts of the entire forum ecosystem considered.
One of the three PUA forums consistently had around 50\% of the monthly active accounts from 2006 until the end of 2017. 
Between late 2017 until mid-2018, we can see a sharp decrease in the activity and in the popularity of \emph{MPUA Forum}.\footnote{We have no explanation for it, but a possible hypothesis is that they had some issues with hosting the standalone website.}
Also, since 2018, the \emph{Rooshv} and \emph{Incels.is} forums have a similar number of active monthly accounts, but the former is much less active (with around 10\% of posts).

On Reddit, from mid-2008 to early 2011, PUA subreddits increasingly grew compared to MRA subreddits, becoming the community with the largest number of monthly active users for a brief period between early 2011 and late 2012 (although they were never more active than their MRA counterparts).
Then in the following years, they decreased in relative activity and popularity, becoming the subreddits with the smallest number of monthly active accounts and posts by the end of 2018.

\descr{Men's Rights Activists.} 
MRA-related communities were the Manosphere pioneers on Reddit, with \emph{/r/MensRights} dating back from March 2008.
The relative number of posts and active accounts decreased over time, but the community was still the most active in number of posts until late 2013.
The absolute number of accounts remained relatively stable since 2013, and the monthly number of posts had a slight decrease. 
However, in relative terms, from 2013 to 2018, the community went from the most popular and active to the second least active, and the third less popular.

In late 2012, we have the first recorded posts in \emph{A Voice for Men}, the largest MRA forum, which is associated with the homonymous blog, running since at least 2009.\footnote{https://web.archive.org/web/*/avoiceformen.com}
The forum experienced a small growth in the number of monthly active accounts and posts in its first year of existence, but then their numbers declined significantly both in absolute terms and in comparison to the other Manosphere communities.

\descr{Incels.}
The first Incel subreddit, \emph{/r/ForeverAlone}, dates back to late 2010, but the community would really pick up momentum in mid-2016. Then, the most active and popular subreddit was not \emph{/r/ForeverAlone}, but \emph{/r/Incels}, created in 2014.
The community would grow to become the most active subreddit in the Manosphere during most of 2017 until \emph{/r/Incels} was banned (in November of 2017~\cite{rincels_ban}). 
According to an Incel-wiki, users of the subreddit then flocked to \emph{/r/Braincels}~\cite{incelwikiBraincels2019} or to \emph{Incels.is}~\cite{incelwikiIncelsCo2019}, a standalone forum created hours after.
Despite the ban, the Incel community on Reddit would continue to prosper, and they became the most active community again in early 2018.
Notice that the ban can be seen in Fig.\ref{fig:all}. 
There is a dip in activity in the Incel-related subreddits, and, around the same time, \emph{Incels.is} was created and already had, in its first month of existence, the highest amount of posts (note how there is a sudden dip in the percentage of posts of other forums after \emph{Incels.is} was created).

\descr{MGTOW.}
The biggest and first MGTOW subreddit \emph{/r/MGTOW} was created in mid-2011, while auxiliary/smaller subreddits such as \emph{/r/MGTOWBooks} were created only later, in 2014. 
Since its creation, the community has steadily grown both in terms of monthly active accounts and monthly posts. 
It was the most active and popular community of the Manosphere for a brief period in late 2017, around the same time when \emph{/r/Incels} was banned, and shortly after became the second most popular community.

In mid-2014, we also have the creation of the \emph{MGTOW Forum}. 
The forum was the most active in 2017, with over 50\% of the total number of posts.
Interestingly, it had around half of the number of unique active accounts compared to \emph{Rooshv}, the second most active forum then.
Around early 2018, the forum decreased both in total number of active accounts and monthly posts, and also in the relative numbers, since \emph{Incels.is }was created around then.

\descr{TRP.} 
Since its creation in October 2012, \emph{/r/TheRedPill} and its adjacent subreddits grew at a fast pace, becoming the most active community by the end of 2013 (in terms of number of posts), and the most popular community by the middle of 2014 (in terms of number of active accounts).
The TRP community would remain the most active community until the end of 2016, when it was surpassed by the Incels community. 
In September 2018, the \emph{/r/TheRedPill} subreddit was quarantined by Reddit, and by the end of the year, it was the community with the third highest number of posts, and the fourth highest number of monthly active accounts.

\descr{Take-aways.} 
Analyzing the popularity and the activity of these forums and subreddits through the years allows us to witness the rise and fall of different online communities.
Importantly, it portrays \textit{when} a community was the most relevant and popular within the Manosphere.
Given the analysis, we can coarsely divide the data-story in three periods: 
a first period (2006-2011) where the Manosphere was mostly about men's rights and pick-up artistry, and where the Incel community was in its infancy,
a second period (2012-2016) where there was a growth of the MGTOW and the TRP community and the decline of PUAs,
and a third period (2016-2018) where Incels and MGTOW rose to prominence, and the TRP-related subreddits lost their momentum.

\begin{figure*}[t]
    \centering
    \includegraphics[width=\textwidth]{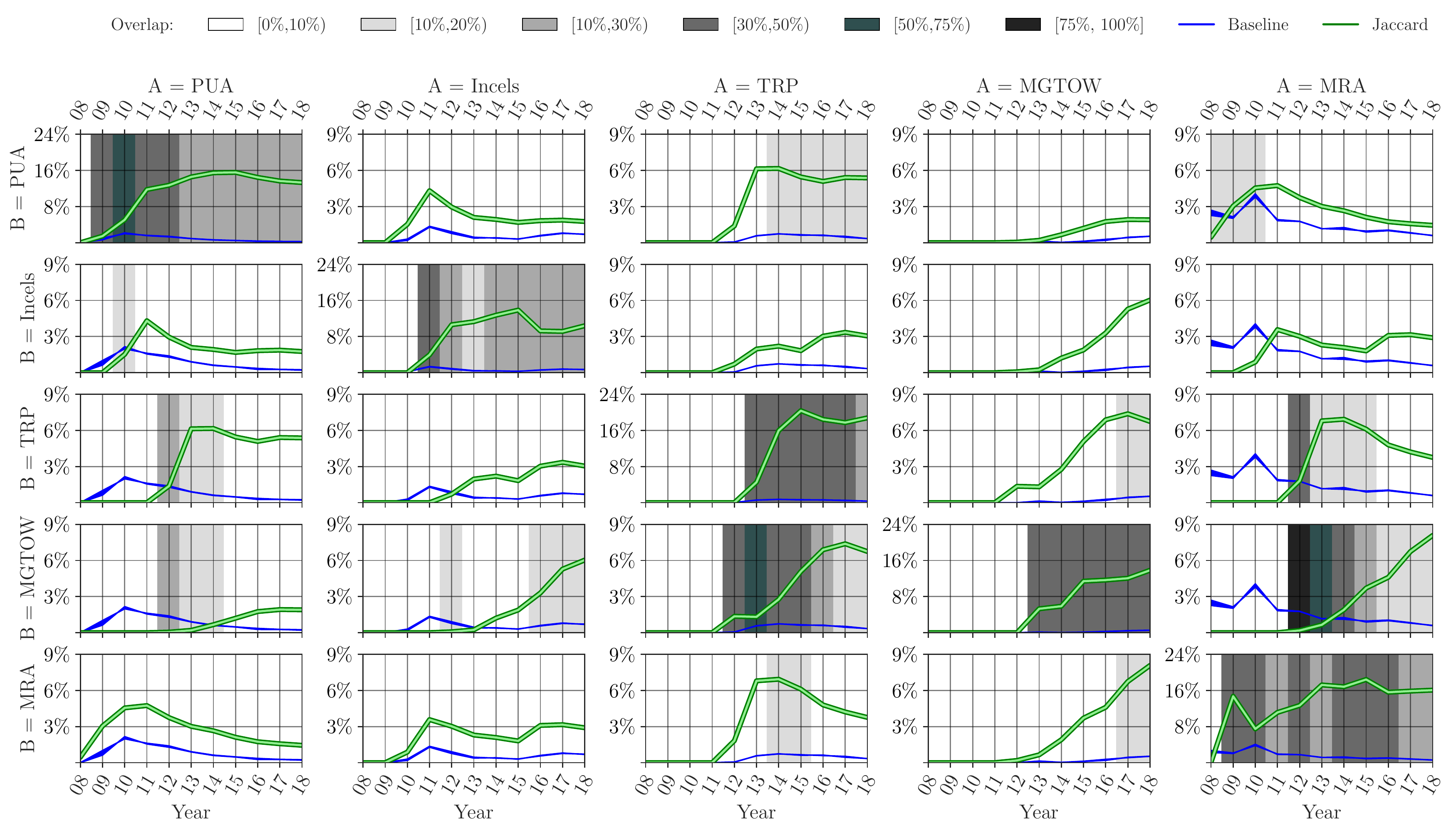}
    \caption{
    We depict the intersection of users in Manosphere communities. 
    The plot reads like a matrix, where each cell shows the similarity between two communities (whose names can be read in the columns and rows).
    The green line in each cell shows the Jaccard similarity between the two communities, and the blue line shows the baseline similarity (as described in the text).
    In the main diagonal, we report the Jaccard similarity and the overlap between a community and itself in the previous year.
    The color on the background of the plot shows the overlap between the two communities.
    The overlap is normalized by the size of the community presented in the rows.
    }
    \label{fig:user_inter_reddit}
\end{figure*}

\begin{figure*}[t]
    \centering
    \includegraphics[width=\textwidth]{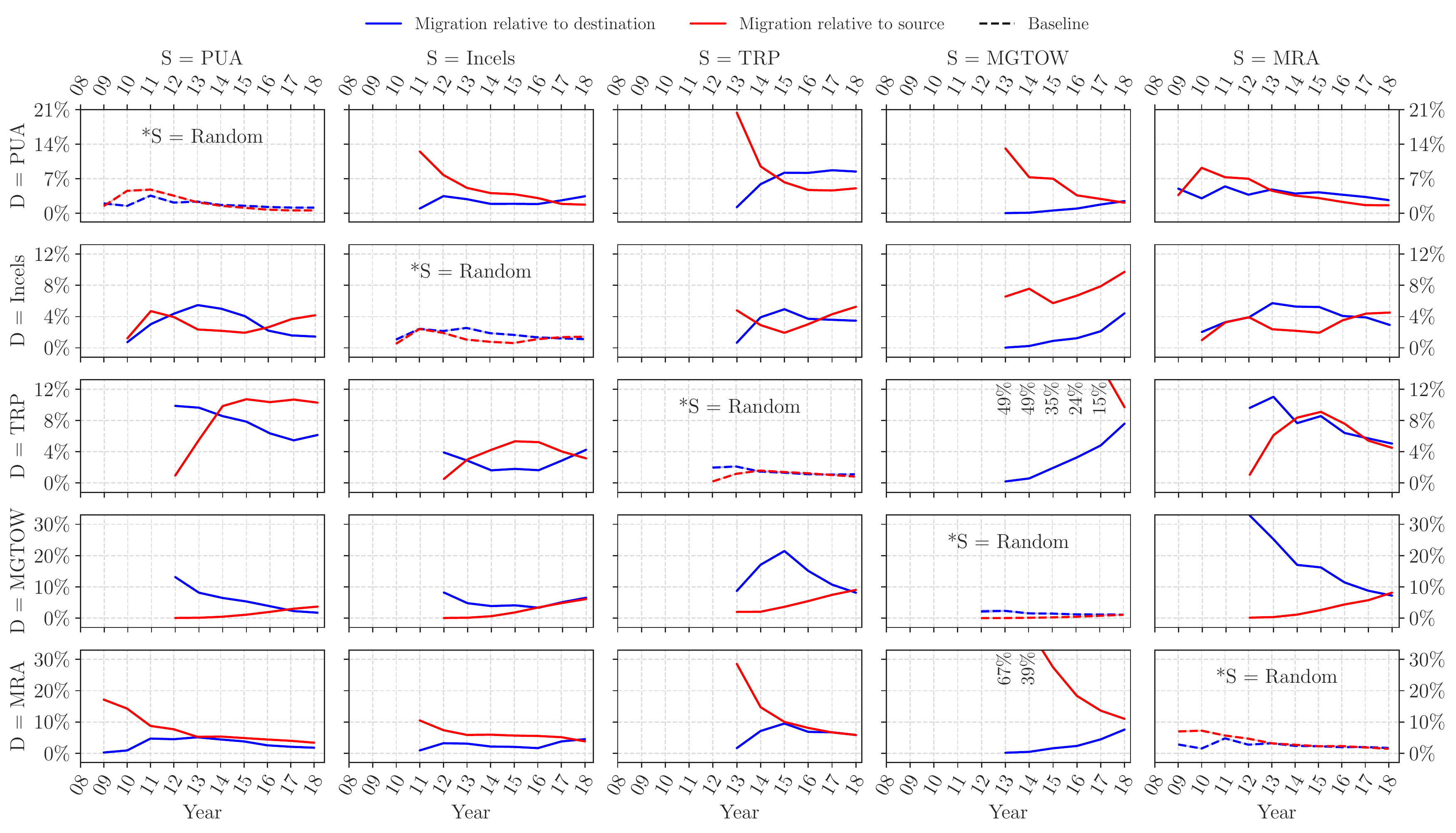}
    \caption{
    Migration of users between the communities of interest.
    Each row has a given community as a target and each column has a community as a source.
    For each given year (x-axis), we report the migration from users who posted in the previous year in the community in the column (the source $S$), to the community in the row (the destination $D$).
    The red line shows the number of users that migrated divided by the original number of users in the source community in the previous year (migration relative to the source).
    The blue line shows the number of users that migrated divided by the number of users in the target community in the given year (migration relative to the destination).
    On the main diagonal, we show the migration relative to the source for the random Reddit sample. 
    95\% CI are drawn for the baseline but are very small.
    }
    \label{fig:user_migra_reddit}
\end{figure*}

\section{User Base Similarity and Migration Across Manosphere Communities (RQ2)}
In this section, we examine the similarities in the user bases of different communities over time, aiming to understand user migration.
We limit our analyses to our Reddit data, as tracking users across forums is hard and inexact.

\descr{User Base Similarity.} 
First, we analyze the intersection between the user bases of different communities using two set similarity metrics: the Jaccard Similarity ($|A \cap B|/|A \cup B|$) and the Overlap Coefficient ($|A \cap B|/|A|$).
The former captures the size of the intersection between sets over the union, while the latter captures the size of the intersection relative to one of the communities.

In Fig.~\ref{fig:user_inter_reddit}, we plot the Jaccard Similarity (lines) and the overlap (background color) between pairs of communities from 2008 to 2018. 
The metrics are computed on a yearly basis, \emph{i.e.}, for a given year we get the set of users in two different communities and calculate the two metrics for them.
Note that, in the figure, the dividing set (in the overlap metric) is the community in the row.
For example, in the cell in the second row and the first column (Incels-PUA), the overlap is calculated with the total number of Incels members as the divisor. 
In the diagonal of the figure, we calculate the similarity metrics of the community with itself the year before (as a point of reference).
Finally, the filled blue interval is a baseline.
To calculate it, we bootstrap $N$ users $1000$ times from the random Reddit baseline and calculate the Jaccard similarity with each of the communities.
We choose $N$ to be of the same size of the community in the column, and thus the baseline is the same for each cell in a given column.
The intuition behind this is that we are calculating the Jaccard similarity with a randomly created community of the same size as the community in the column.
This is a lower bound, as in, if the similarity between two of the communities was smaller than the similarity of one community with the random baseline, then it is unlikely to be noteworthy.

The Jaccard similarity between the MRA and the MGTOW subreddits shows a sharp increasing trend, reaching close to $9\%$ in 2018.
This number is higher than the self-similarity of the MGTOW community in its founding years (2012--2014).
Looking at the overlap, we can see that MGTOW was almost a subset of the MRA in 2012 ($87\%$ overlap) and 2013 ($52\%$ overlap).
The MRA community was also quite similar to the PUA community around 2010. 
In this period, it had an overlap of over 10\% with respect to the PUA community and a Jaccard Similarity of around 4.5\%. 
Notice that, in 2010, the Jaccard Similarity of the PUA community itself in the previous year was around 6\%.

The TRP community exhibits high similarity with several other communities throughout the years. 
In its creation, it had a high overlap with the PUA ($25\%$ in 2012) and MRA communities ($33\%$ also in 2012), and in subsequent years its Jaccard Similarity with these two communities reached around $6\%$. 
In the early years of MGTOW, it also had substantial overlap with the community (more than $35\%$ between 2012 and 2015). 
In more recent years (between 2016 to 2018), the Jaccard similarity with the MGTOW user base is over $6\%$ and the overlap with regards to TRP is over $10\%$.
In fact, $25\%$ of the users who commented in TRP subreddits during their first year of existence (2012) were also active in PUA forums (438 users), and out of the initial users in the early days of MGTOW subreddits (2012--2014), $11\%$ also commented in PUA forums in the same year (266 users).

The Incels community had more than $10\%$ overlap with the PUA community back in 2010; recall that the PUA community was much bigger back then with $1,367$ vs. $8,632$ active users. 
Also, after 2016 both the similarity and overlap with the MGTOW community increased, with the former being of around 6\% (in comparison, the self-similarity of the Incels subreddit after 2016 was only around 10\%), and the latter of around 10\% (with respect to the MGTOW user base). 
Of all communities that have grown in popularity in most recent years as discussed in the previous section, the Incel community had the least overlap with other communities in its founding years.

This analysis shows that Manosphere communities shared a substantial portion of their user base, especially in the founding years of newer communities, such as TRP or MGTOW.
This effect is less noticeable for the Incel community, although it was similar to the PUA community in its early years, and recently, increasingly shares its user base with the MGTOW community.

\descr{User Migration.}
To better understand how users have navigated between the communities in the Manosphere, we perform a different analysis.
For each pair of communities ($S$ and $D$) and each pair of subsequent years ($i-1$ and $i$), we track the number of migrants: users that posted in a ``source'' community ($S$) in a given year ($i-1$) and then proceeded to post in ``destination'' community ($D$) in the subsequent year ($i$).
We define the \textit{migration relative to the destination} (MRD) as the number of migrant users divided by the size of the ``destination'' community $D$ in time $i$, and \textit{migration relative to the source} (MRS) as the number of migrant users divided by the size of the ``source'' community $S$ in time $i-1$.
Suppose for instance that 10 users commented in $D$ at year $i$, out of which 6 have commented in $S$ in year $i-1$; this means MRD is $6/10$, \textit{i.e.}, out of the 10 users in $D$, we can trace 6 back to $S$. 
Now assume that, in community $S$, there are 12 users in year $i-1$: MRS is $6/12$, \textit{i.e.}, half the users from community $S$ went on to comment in community $D$ in the subsequent year.
It is important to emphasize that users can be active in multiple communities and that these metrics, by themselves, may be more analogous to a spread than a migration. 
To help with that, we analyze them along with the timeline of the popularity of/activity in the communities, depicted in Fig.~\ref{fig:all}.

To obtain a baseline for this scenario, we once again use the Reddit $0.5\%$ random sample. 
We calculate MRS and MRD with one of the communities as the destination $D$, and with the random sample as source $S$.
Since the random sample is orders of magnitude larger than the communities, we under-sample it and draw $k$ random users. 
We choose $k$ to be equal to the size of the largest Manosphere community in the year of interest.
Suppose that we are calculating the migration from this random baseline to Incels in 2012.
We first get the users who commented in Incel subreddits in 2012.
Next, we get the size $k$ of the biggest community one year before (2011).
Then we sample $k$ users who commented in 2011 from the random subreddit sample and compare MRS and MRD using these two subsets.
We repeat the procedure 100 times per community and report 95\% CIs.

Results for this analysis are shown in Fig.~\ref{fig:user_migra_reddit}. 
Each column represents a different community as the source $S$, and each row, a destination community $D$. 
The main diagonal contains the baseline, where the source community is the random sample.
Importantly, in our analysis of the migratory fluxes, we must keep in mind the order in which the communities were created and rose into popularity (c.f. Fig.~\ref{fig:all}).

First, we find that the MRA to MGTOW migration did indeed happen. 
This can be seen by inspecting the MRD with MRA as a source and MGTOW as the destination.
In the early years of MGTOW, $>30\%$ (2012) and $>20\%$ (2013) of the commenting users had commented in MRA communities in the year before.
Interestingly, as the MGTOW community grew, the migration relative to destination (MRD) became less prevalent, and the migration relative to source (MRS) became more prevalent.
The MRA community also showed substantial migrations to the TRP community. 
Both in 2012 and 2013, the founding years of the TRP subreddits, we have around 10\% MRD from the MRA community.
Moreover, we see that the MRS grew as the TRP community rose to prominence (see Fig.~\ref{fig:all}), reaching around 10\% in 2015.

We also find that the PUA community had substantial migration to other communities.
Examining MRS and the MRD of the first column, where PUA is the migration source, we find that the MRD towards TRP and MGTOW subreddits was around $10\%$ in 2012: approximately 1 out of 10 users of each of these communities in 2012 had previously (in 2011) posted in PUA subreddits.
Also, there was substantial migration to MRA subreddits in the early days of PUAs on Reddit (2009-2011).
For instance, in both 2009 and 2010, there was $>10\%$ MRS. 
Recall that back then the MRA community on Reddit was much bigger than the PUA.

Users of the PUA community also migrated to the Incel subreddits. 
The migration reached a peak of around 4\% of MRS in 2011 of around 5\% MRD two years later, in 2013. Although not so prominent, this was still twice the random sample baseline, which had around 2\% MRS and MRD on both these occasions.
The reverse migratory flow (from Incel to PUA) was more substantial.
In 2010, we observe an MRS of 13\% and in the following year, of 7\%.

Other interesting migrations are those between MGTOW and TRP, and Incels and TRP.
Recall that these communities gained traction more recently. 
We find that in the early years of the TRP community, when it was much smaller than MGTOW, there was substantial migration of the small MGTOW user base to the TRP user base, with around 49\% MRS in 2013 and 2014.
Also, in the early years of the MGTOW community, in 2012 and 2013, when the Incel community was much bigger, there was over 10\% MRD from the Incel community to the MGTOW community.

\descr{Take-aways.} Overall, dissecting these migratory influxes is difficult, and is a fruitful direction for future work. 
Yet, our analyses indicate that there are substantial user overlaps in the Manosphere communities throughout history, including some large migratory movements from older communities such as MRAs and PUAs to newer ones, such as TRP, MGTOW, and Incels.
This suggests that it is inaccurate to think of the Manosphere simply as discrete communities that experience peaks in popularity and then decay.
It is important to consider that users were active in different communities simultaneously and that some communities are jump-started by large migratory influxes of older existing ones.

\begin{figure*}[t]
\center
\includegraphics[width=\textwidth]{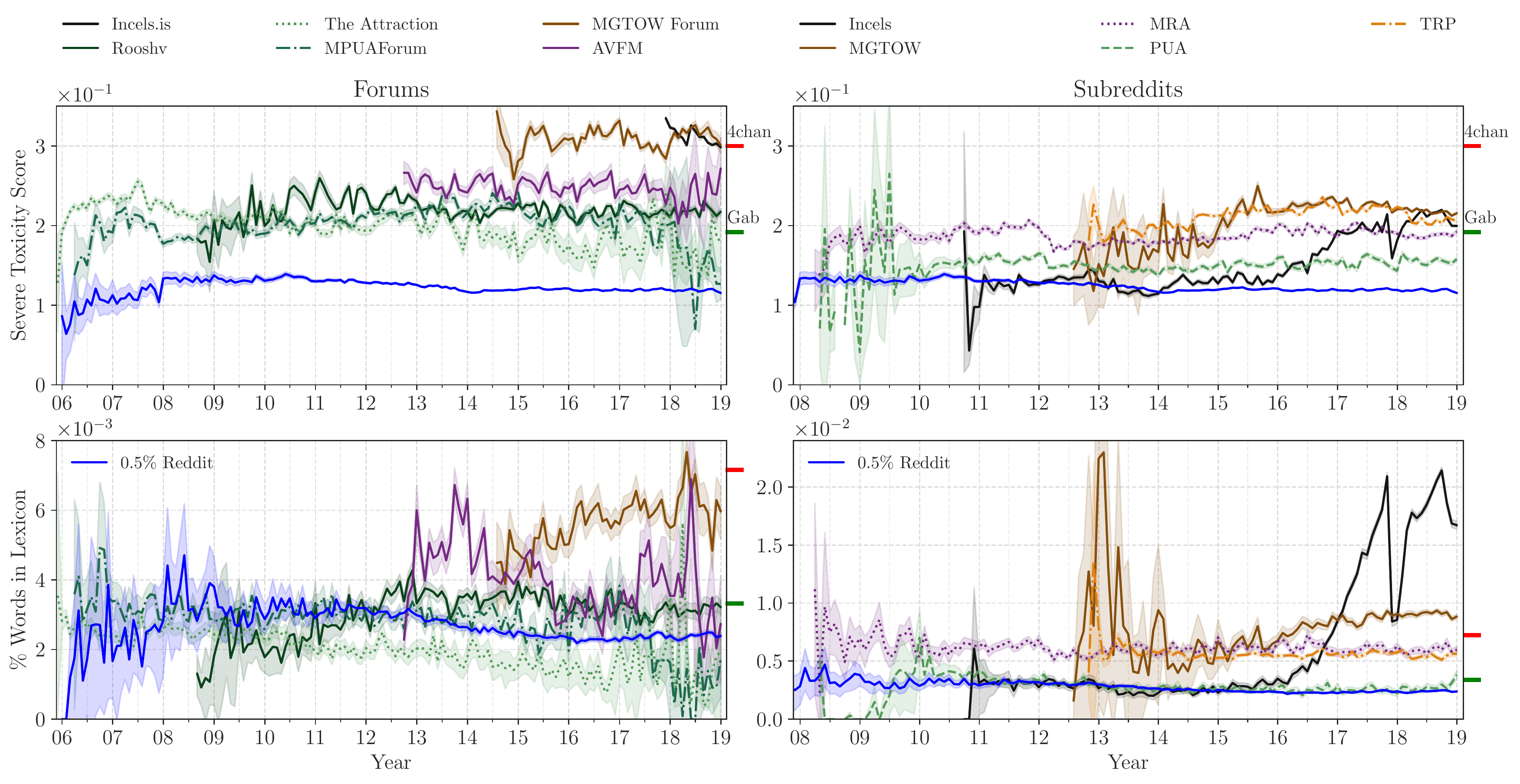}
\caption{Evolution of toxic and misogynistic speech in the Manosphere using Perspective API (top row) and~\cite{farrell2019exploring}'s lexicon (bottom row). 
We report mean toxicity score/lexicon occurrence, each month along with 95\% bootstrapped confidence intervals.
Note that the toxicity scores for the baseline datasets are in all plots: the random Reddit dataset is the vibrant blue dashed line, and the Gab and 4chan datasets are on the right-side of each plot, as two small rectangles (red for 4chan and green for Gab) representing the 95\% CI of their means.}
\label{fig:all_toxicity}
\end{figure*}

\section{Evolution of Toxicity and Misogyny\\in the Manosphere (RQ3)}

We now study the spread of toxicity and misogyny in the Manosphere over time, and how it compares to other mainstream and fringe Web communities.

\descr{Meaasuring Toxicity.} As a proxy for toxicity, we rely on the \emph{Severe Toxicity} score returned by the Perspective API~\cite{jigsaw2018perspective}; more precisely we run all posts from all forums as well as all posts from all subreddits through the API and obtain the severe toxicity score for each post.
To calculate the severe toxicity score, Perspective uses a Convolutional Neural Network (CNN) trained with GloVe word embeddings, which are fine-tuned during training.
The severe toxicity score ranges from 0 to 1, and it is defined to be high for ``a very hateful, aggressive, disrespectful comment or otherwise very likely to make a user leave a discussion or give up on sharing their perspective.''
Compared to the normal Toxicity score, it relies on a model that is less sensitive to positive uses of profanity.

\descr{Measuring Misogyny.} As a proxy for misogyny, we use a modified version of the misogyny lexicon from~\cite{farrell2019exploring}.
The lexicon was designed to capture different aspects of misogynistic speech online.
We first remove from the lexicon the names of the communities themselves (e.g., ``mra'', ``mgtow'') and do not use three out of the nine categories in the lexicon (``Racism'', ``Homophobia'' and ``Physical Violence'') since they are composed of more common inappropriate words (e.g., ``f*g'', ``n*gger'', ``shoot''), which expose the link to more general expressions of hate but also add some noise to the more focused expression of misogyny that the Manosphere ``specializes'' in.
These are better represented in other categories such as ``Patriarchy'' and ``Sexual Violence,'' with words such as ``betabux'', ``stacy'', ``cock carousel.''~\footnote{For
more context on these terms, see \url{rationalwiki.org/wiki/Manosphere_glossary}.}
Using the lexicon, we count exact matches for each post in all subreddits, forums, and baseline datasets (recall that baseline datasets are large samples from Reddit, Gab, and 4chan used as reference points for our analyses). 
We then normalize them by the total number of words in that post and call it \emph{lexicon density}.

\descr{Overall evolution.} Fig.~\ref{fig:all_toxicity} reports the evolution of toxicity and misogyny in the Manosphere according to both Perspective API and \cite{farrell2019exploring}'s lexicon.
For two of the baseline datasets (Gab and 4chan), the data is very recent (2016 onwards), and thus we measure the mean toxicity/lexicon density as a small rectangle on the side of the plot.
Notice that we opt to report their means like this because there is little variation in their toxicity score,\footnote{In 4chan, for the toxicity score, min=$0.282$, max=$0.316$, and for the lexicon density, min=$0.00655$, max=$0.00812$. For Gab, min=$0.171$, max=$0.208$, and min=$0.00240$, max=$0.00367$} and because its confidence interval is very tight ($<10^{-4}$), due to the size of the datasets. 

\descr{Toxicity and Misogyny within communities.}
Looking at the evolution within each community, a noteworthy trend is a sharp rise in the mean toxicity score and lexicon density for both MGTOWs and Incels.
For MGTOWs, there are few posts in the early years (see Fig.~\ref{fig:all}) thus yielding larger CI, but there is an overall positive trend from 2013 to late 2016 for toxicity, and from 2015 to 2019 with respect to lexicon density.
For Incels, the sharp trend starts in 2016, which is also when the community gained a lot of traction in terms of users and number of posts (again, see Fig.~\ref{fig:all}). 
As of 2019, we find that both these communities present levels of toxicity similar to Gab, and lexicon density comparable to either Gab (for Incels) or 4chan (for MGTOW).

Within single communities, almost all forums are significantly more toxic than their subreddit counterparts---except for MRA subreddits and forums, which exhibit similar levels. 
For example, PUA's mean toxicity in the subreddits is  around 0.15, whereas, for the PUA-related forums, such as \emph{The Attraction} and \emph{MPUA Forum}, the mean toxicity score is consistently around 0.20 for most of the period the forums were active. 
This same trend is noticeable when comparing Incels subreddit and the \emph{Incels.is} forum, and the MGTOW subreddits and the \emph{MGTOW Forum}.

Analyzing the toxicity on the Incel forum is particularly interesting as it is possible that the community was initially entirely made up of old \emph{/r/Incels} members~\cite{incelwikiIncelsCo2019}.
By comparing the toxicity of Incel subreddits  when \emph{/r/Incels} got banned, in November 2017, with the initial toxicity of \emph{Incels.is}, we indeed see a rise in toxicity; from 0.2 to 0.3 on average. 
In terms of the baseline communities, this is more or less a jump from the level of toxicity in Gab to the level in 4chan.
Although outside the scope of this paper, this finding highlights the potential pitfalls of de-platforming measures taken by mainstream platforms: while the banning subreddits might indeed reduce toxicity levels across Reddit~\cite{chandrasekharan2017you}, users migrating to less moderated platforms might become substantially \emph{more} toxic.
Future work exploring this direction would need to more carefully inspect confounding variables, such as the length of the posts (which may differ between the forum and the subreddit and be correlated with toxicity).

However, the pattern of forums being more toxic than subreddits does not appear with the misogyny lexicon-based analysis. 
There, lexicon density is similar for the same community in forums and subreddits.
This result somewhat highlights the limitations of the lexicon-based approach.
For example, it yields similar values for Reddit and \emph{Rooshv}, which by 2012 was already listed by the Southern Poverty Law Center as a notoriously misogynistic website~\cite{shugermanNoxiousPickupArtist2019}.
Despite these limitations, the lexicon-based analysis does reveal similar patterns in the toxicity rise for the MGTOW and the Incel subreddits.

\descr{Comparing communities.}
When comparing the different communities, we find that, for both scenarios (forums and subreddits) and measurements (toxicity and misogyny), Incels and MGTOW have appreciably higher scores than the remaining communities. 
For the TRP community, which exists only on Reddit, results are mixed.
Considering only the toxicity score, the TRP community yields similar toxicity scores as the MGTOW community, yet in the lexicon analysis, their scores are similar to those of MRAs (which are substantially less toxic).
Overall, we can rank communities based on the levels of toxicity and misogyny, with MGTOW, Incels and TRP at the top, MRAs in the middle, and PUAs at the bottom.

\descr{Remarks.} Naturally, comparing the communities at face value is challenging because stylistic norms and communities' standards may differ, and is unclear how these may affect our measurement tools.
However, our analysis uses different measurements and compares the same communities across time and on different platforms.
Our findings should be considered in light of the results presented earlier.
For example, many of the individuals of the MGTOW and the TRP communities can be traced, respectively, from the MRA and the PUA communities (see Fig.~\ref{fig:user_migra_reddit}).
In that sense, the ``evolution'' of the Manosphere communities coincides with a substantial increase in toxicity and these results confirm, at scale, qualitative work that has theorized a worrying path of radicalization.

Furthermore, the two most toxic/misogynistic communities have been subject to global moderation actions by Reddit.
The biggest Incel subreddit, \emph{/r/Braincels}, was banned in September 2019,
and the biggest MGTOW one, \emph{/r/MGTOW}, quarantined in January 2020,
whereas, less toxic/misogynistic subreddits, e.g., \emph{/r/seduction} and \emph{/r/MensRights}, remain active as of May 2020.

\section{Discussion and Conclusion}

This paper presented a data-driven characterization of the Manosphere across the Web over the last 14 years.
We gathered and analyzed a large dataset including  dedicated forums and subreddits,
and used a typology to divide these into communities.
Our analysis provides a partial reconstruction of the Manosphere's history, along with how its language evolved in terms of toxicity and misogyny.

By analyzing the popularity and user activity across several different forums and subreddits, we find that older communities, such as Men's Rights Activists and Pick Up Artists, are becoming less popular and active, while newer communities, like Incels and Men Going Their Own Way, are thriving.
Besides simply looking at the activity in each community, we also examined (on Reddit) the user overlap and migratory fluxes.
We find that communities in the Manosphere shared large amounts of users throughout their history and that there was substantial migration from older to newer communities. 
For example, there was a migratory influx of MRA users to MGTOW subreddits, with over 50\% overlap in the first two years of its existence.
Our analysis unveils the magnitude of links between these communities, but also that it is important to consider the Manosphere as a more cohesive whole: some users participate in several different communities simultaneously and the user base of some communities can be directly traced back to previous older ones.

Our analysis indicates that the newer communities in the Manosphere (Incels, MGTOW, and TRP) are more toxic and misogynistic than the older ones (PUA, MRA), as theorized by previous literature.
It is important to stress, however, that this analysis has its limitations. 
As discussed, we use the Perspective API, which previous work shows that can be deceived~\cite{hosseini2017deceiving} and may exhibit racial biases~\cite{sap2019risk}.
Also, it is reasonable to assume that differences in Perspective scores may arise from varying use of the English language and the varying perception of toxicity across geographies.
We try to minimize this problem by comparing the communities across time, platforms, and by using a misogyny lexicon~\cite{farrell2019exploring}'s.

It is also important to consider our work in the broader context of the existing scholarship on the Manosphere.
Our study not only provides large-scale quantitative analysis that supports much of the core hypotheses of qualitative work but also prompts new research directions.
The migration from MRA to MGTOW corroborates Lilly's work~\cite{lillyWorldNotSafe2016}, in that both communities are similar with respect to their views on the root of the alleged crisis of masculinity and femininization of society.
Yet, the increasing migration from MGTOW towards Incels -- who, according to Lilly, see the crisis of masculinity in the femininization of men -- still lacks theoretical explanation: are these users changing their minds? Or are Incels increasingly seeing the crisis of masculinity as a societal issue as opposed to a personal one?
Moreover,~\citet{gingAlphasBetasIncels2017} states that content produced by the Manosphere, despite internal conflicts and contradiction, is united in its adherence to ``Red Pill'' concepts.
In our work, we find evidence, given significant overlap and migration across communities, that these contradictory ideas may very well be steps along a radicalization pathway.
Many of the individuals involved with the PUA community went on to participate in more extreme anti-feminist communities such as TRP, whose users, in turn, migrated to MGTOW.
In this sense, further work is needed to understand the underlying processes that lead to these shifts in views, and the Manosphere can serve as a fruitful area of exploration considering its long history of spawning new, seemingly more extreme sub-communities.


Lastly, it would also be interesting to analyze migrations between the user bases of Manosphere communities to other fringe communities.
For instance, perpetrators of hate crimes were associated with both White Nationalist ideology and Manosphere-related ideas~\cite{hanau_2020,timothy_2017}, therefore, it is possible that communities in the Manosphere may act as real gateways along a radicalization pathway.

\section{Acknowledgements}

B.B. was partially supported by NSF under grant DMR-1945058. G.S. was partially supported by NSF under grant CNS-1942610. E.D.C. was partially supported by The National Research Centre on Privacy, Harm Reduction and Adversarial Influence Online (UKRI grant: EP/V011189/1).

{
\small
\bibliography{ref}
 }

\end{document}